\documentclass{acm_proc_article-sp}

\makeatletter
\newif\if@restonecol
\makeatother

\usepackage[linesnumbered, ruled]{algorithm2e}
\usepackage{graphicx}
\usepackage{comment}
\usepackage{amsmath}
\usepackage{ctable}
\usepackage{balance}
\newcommand{\sig}{$SiG$}
\newcommand{\A}{$\mathcal{J}$}
\newtheorem{thm}{Theorem}
\newtheorem{definition}{Definition} 
\begin{document}

\title{Silence is Golden}
\subtitle{exploiting jamming and radio silence to communicate
\titlenote{This paper was (unsuccesfully) submitted to ACM CCS 2013.}}
\numberofauthors{2}
\author{
\alignauthor
Roberto Di Pietro \\
       \affaddr{Dipartimento di Matematica e Fisica}\\
       \affaddr{Universit\`{a} di Roma Tre}\\
       \affaddr{Rome, Italy}\\
       \email{dipietro@mat.uniroma3.it}
       \alignauthor
       Gabriele Oligeri \\
       \affaddr{Dipartimento di Ingegneria e Scienza dell'Informazione}\\
       \affaddr{Universit\`{a} di Trento}\\
       \affaddr{Trento, Italy}\\
       \email{gabriele.oligeri@gmail.com}
} 
\maketitle
\begin{abstract}
  Jamming techniques require just moderate resources to be deployed, while their
  effectiveness in disrupting communications is unprecedented.  In this paper we introduce
  several contributions to jamming mitigation.  In particular, we introduce a novel
  adversary model that has both (unlimited) jamming reactive capabilities as well as
  powerful (but limited) proactive jamming capabilities.  Under this powerful but yet
  realistic adversary model, the communication bandwidth provided by current anti-jamming
  solutions drops to zero.
  \\
  We then present Silence is Golden (\sig): a novel anti jamming protocol that,
  introducing a tunable, asymmetric communication channel, is able to mitigate the
  adversary capabilities, enabling the parties to communicate. For instance, with \sig\ it
  is possible to deliver a 128 bits long message with a probability greater than 99\% in
  4096 time slots in the presence of a jammer that jams all the on-the-fly communications
  and the 74\% of the silent radio spectrum---while competing proposals simply fail.
  \\
  The provided solution enjoys a thorough theoretical analysis and is supported by
  extensive experimental results, showing the viability of our proposal.
\end{abstract}
\section{Introduction}
\label{sec:introduction}
Wireless communications are prone to several kinds of attacks due to the shared nature of
the radio channel. \emph{Jamming} is one of the most effective denial of service attack
that might be performed in such a scenario \cite{poisel}. Jamming is a general term that
refers to several disruptive radio activities 
aiming at either to interfere or to prevent communications.  While jamming originated in
the military scenario, it is nowadays a threat also to civilian communications
\cite{richa}.  There are mainly two reasons for the widespread diffusion of jamming as a
DoS attack in the wireless scenario: the first one is its effectiveness; the other one is
that its implementation does not require specialized hardware. For instance, a cheap WiFi
radio can be used to generate collisions with the on-the-fly packets so that the receiver
cannot decode them \cite{dehghan}; or, the same radio might be used to occupy the
transmission channel in such a way that the transmitter cannot even start a new
communication \cite{thuente}.  In the last decade, jamming devices have been evolved into
high power random noise transmitters, making the activity of jamming a dreadful threat for
wireless communications. As an example, we observe that military equipments implement band
jamming by transmitting a white noise signal of 100W power over several frequency bands
between 20 and 2483 Mhz \cite{phantom}.
\\
Different jamming strategies have been deployed during the years \cite{networkCR13}.  The
jammer might target the transmitter side by performing the so called noise-spoofing, i.e.,
the transmitter never senses a clear radio channel and therefore never starts a new
transmission. Conversely, the jammer might target the receiver side, i.e., noise
jamming. Noise jamming interferes with the current on-the-fly message introducing several
errors or even making it not receivable.
Different radio techniques are possible in order to interfere with the transmitter and the
receiver radios \cite{poisel}: the jammer might perform a tone jamming by generating a
sinusoidal waveform whose power is concentrated on the target carrier frequency or it
might perform a band jamming by spreading a flat spectrum power in the bandwidth of
interest.
\\
Jammers can be categorized in two main families: \emph{proactive} and \emph{reactive}. The
proactive jammer randomly jams $A$ of the $F$ available frequencies in the radio spectrum,
and therefore, the transmitter/receiver pair has only $F-A$ frequency bands in order
to communicate. The solutions dealing with the proactive jammer implement a strategy based
on a trial-and-fail communication process in random frequency slots. Conversely, the
reactive jammer is more effective, in fact, it jams the communication after it has sensed
it. The reactive jammer senses the radio spectrum and jams the communications as soon as
they appear on-the-air. Dealing with a reactive jammer is more difficult: the
sense-and-jam behavior is always disruptive, and so far, all the proposed solutions
leverage a bounded jammer model, e.g., the size of the jammed area, the reaction time, the
number of the jammed frequencies.
\\
{\bf Contribution.} In this work we provide several contributions: first, we introduce a
novel type of adversary.  Our adversary combines both \emph{reactive} and \emph{proactive}
jamming capabilities.  As a reactive jammer, it is able to disrupt all the on-the-air
communications in the available spectrum.  In fact, in this capacity, we assume it is able
to perform both a network and a spectrum wide sensing and to react with a jamming signal
emitted without delay with respect to the transmission sender.  As a proactive jammer, our
adversary is able to jam additional (apparently) not used frequency slots in the radio
spectrum. To the best of our knowledge, these are the most strongest assumptions ever made
in the literature for a radio jammer.
The second contribution is the definition of the \sig\ protocol.  Our proposal is, to the
best of our knowledge, the only one able to thwart the powerful, yet viable, adversary
above introduced.  For instance, against the introduced adversary, the \sig\ protocol
enables the transmitter to successfully deliver a 128 bits long message within 4096 time
slots while other solutions in the literature, cannot guarantee any communications at all.
\\
Further, the \sig\ protocol is fully detailed, showing how to tune its parameters in order
to trade-off communication capabilities with an efficient usage of the transmission slots.
Moreover, a thorough analysis of its capabilities is provided, together with a comparison
against anti-jamming state of the art solutions.  Finally, an extensive simulation
campaign supports our findings.
 
Section \ref{sec:scenario} introduces our reference scenario, defining the transmitter,
the receiver, and the adversary model.  Section \ref{sec:first-look-at} presents a
simplified version of the \sig\ protocol, which is subsequently detailed in Sections
\ref{sec:enabl-freq-hopp} and \ref{sec:binary-asymm-error}, that introduce the frequency
hopping scheme and the error correcting codes, respectively. Section \ref{sec:protocol}
provides a detailed decryption of the \sig\ protocol and Section \ref{sec:perf-eval} shows
the performance of \sig\ by means of a theoretical analysis and simulation
results. Finally, Section \ref{sec:comp-with-other} compares \sig\ with other recent
solutions to jamming attacks. Some concluding remarks are reported in Section
\ref{sec:conclusions}.
\section{Related work}
\label{sec:related-work}
In the following, we review the most relevant anti-jamming techniques. We consider both
proactive and reactive anti-jamming techniques, and finally, we recall an early solution
that leverages jamming to communicate between peers.
\subsection{Proactive jamming}
\label{sec:proactive-jamming}
An early analysis on the feasibility of launching and detecting jamming attacks in
wireless networks is proposed in \cite{trappe}. Authors provide an in-deep study about the
problem of conducting radio interference attacks on wireless networks, and examine the
critical issue of diagnosing the presence of jamming attacks. They consider different
adversarial models and run real test-beds to measure the adversarial performance. They
show that by using signal strength, carrier sensing time, or the packet delivery ratio
individually, it might difficult to conclude the presence of a jammer.
\\
Many solutions have been proposed against proactive jammers. We identify two main
families: the ``keyed'' \cite{lazoswisec} and the ``key-less'' \cite{capkun_jsac}. The
former leverages a pre-shared secret in order to generate frequency hopping sequences
(unknown to the jammer), while the latter leverages a delay between the sender and
receivers in order to make them converging on a shared transmission frequency. Authors in
\cite{lazostmc}\cite{lazoswisec} propose the Time Delayed Broadcast Scheme (TDBS): a
broadcast communication is achieved by means of a sequence of unicast
communications---sometimes assisted by proxies.  The solution relies on long frequency
hopping sequences that are pre-loaded in each sensor belonging to the network before nodes
deployment. A key-less solution is presented in \cite{capkun_jsac}. Authors propose to
deliver a message between two peers by an uncoordinated spread spectrum technique while
introducing a delay between the transmitter and the receiver in order to have them
synched. Finally, an early key-less solution is from \cite{baird}. Authors leveraged
specialized ultra-wide band radios in order to transmit short impulses. Such communication
scheme is difficult to jam, i.e., so far radio impulse cannot be cancelled with an inverse
waveform. Each bit of the message is coded with a time-delayed radio impulse, nevertheless
spurious impulses (errors) might appear at the receiver side due to noise fluctuations or
malicious entities, and the previous produces an enormous increase in the computational
cost that is exponential in the size of the message.
\subsection{Reactive jamming}
\label{sec:reactive-jamming}
Reactive jamming involves the activity of sensing the channel and subsequently switching
the radio to the jamming status.  As for reactive-jamming, the current state of the art
solutions do not deal directly with the jammer but leverage either space or temporal
bounds the adversary is subject to.
Authors in \cite{liu} propose to exploit the reaction time of the reactive jammer in order
to enable communication; they argue that the jamming activity needs more than $t_s = 1 ms$
while radio switching needs other $t_c = 50 \mu s$, while the transmitter has already sent
$R (t_s+t_c)$ bits---assuming a transmission rate $R$. The receiver collects all the bits
that are transmitted by the sender but not jammed by the reactive jammer, and assembles
them to construct the original message.
\\
In \cite{xuan}, author propose a combined solution that involves both locating the
reactive jammer and deactivating the nodes that trigger its activity. Authors observe that
the reactive jamming activity is particularly disruptive in dense WSNs. Indeed, the
reactive jammer is triggered by a specific node, while the jamming signal will eventually
prevent all the communications of the nodes in the jammer neighborhood. In order to avoid
this, authors propose a solution where nodes cooperate in order to estimate the jammer
position, and subsequently enforce the radio-silence of the nodes that trigger the jamming
activity.
\\
Another solution to reactive jamming is POWJAM \cite{hamieh}. Author proposes
short-distance transmissions (with low power) between peers in order to hidden the
transmitter to the reactive jammer. Each long-range communication turns out to be
implemented by a sequence of multi-hop transmissions characterized by low-distance
propagation and therefore a low probability to be sensed by the jammer.
\\
An efficient and fair MAC protocol robust to reactive interference has been proposed in
\cite{awerbuch} and subsequently extended in \cite{richa}. The proposed protocol is robust
to both internal and external interference requiring no knowledge about the number of
participants, nevertheless the authors bound the reactive jamming activity to
$(1-\epsilon)$-portion of the available time slots.
\\
Another interesting solution comes from \cite{blass}. There, authors design, prototype,
and evaluate a system for cancelling the jamming signal: the system combines a mechanical
beam-forming design with an auto-configuration algorithm and a software radio digital
interference cancellation algorithm. The mechanical beam-forming uses a custom-designed
two-elements antenna architecture and an iterative algorithm for jammer signal
identification and cancellation.
\\
Finally, an interesting solution is provided in \cite{bhadra}. The scenario involves 4
nodes and a slotted channel: two legitimate peers communicate to each others by
transmitting messages on the shared channel, while one of the illegitimate users
interferes/jams the legitimate messages. Now, the other illegitimate peer decides a
reception of a ``1'' when a collision is detected, while decides for ``0'' when the slot
is empty or filled up by a legitimate message. Authors prove that the status of the
channel, i.e., jammed or not-jammed, can be used to communicate one bit of information.
\section{Scenario}
\label{sec:scenario}
We consider a wireless communication scenario constituted by a point-to-point link between
a transmitter ($T$) and a receiver ($R$), where $T$ wants to deliver a message $m$
constituted by $L_{m}$ bits to $R$. We assume $m \in \Phi$, where $\Phi$ is a dictionary
shared between $T$ and $R$ (note that such a dictionary could be the set of correct
English words).  Further, we assume the radio spectrum as constituted by $F$ different
frequency bands (channels), i.e., $\{f_0, \ldots, f_{F-1}\}$.  Both the transmitter and
the receiver share a pre-loaded secret $s_0$.
\\
\\
Generally speaking, $R$ could receive from $T$ ---due to the jamming activity and the
radio noise--- a message $m'$, such that $m' \neq m$. The \sig\ protocol guarantees (with
a given, tunable, probability) to recover $m$.  In the following, we assume $m$ to be
constituted by a few bits, e.g., $L_m \in \{128, 256, 512 \}$ bits---e.g. $m$ could carry
commands or geographical coordinates.
\\
\\
Further, we assume that $T$ and $R$ are loosely time synchronized \cite{ganeriwal} and
that time is divided into slots, i.e., $i \in [0, \ldots, \infty[$. Finally, we do not
assume any specific or powerful hardware configuration at both $T$ and $R$ --- the \sig\
protocol only needs the computation of a cryptographically secure hash function
\cite{sha1} and the capability to run a symmetric encryption algorithm such as AES
\cite{aes}.
\\
\\
Table \ref{symbols} shows a resume of the symbols and acronyms used throughout this paper.
{\footnotesize
\begin{table}[t]
\caption{Notation summary.\label{symbols}}
\begin{tabular}{l|l}
  $T$, $R$, \A                               & transmitter, receiver, and jammer \\
  $m$, $m'$                               & transmitted and received messages  \\
  $\Phi$                                  & message dictionary \\     
  $m_e$, $m'_e$                             & transmitted and received encrypted messages \\
  $L_m$                            & length of the messages $m$, $m_e$, $m'$, and $m'_e$\\ 
  $m_{ec}$, $m'_{ec}$                         & transmitted and received \\
  & encoded encrypted messages \\
  $L_c$                             & length of the messages $m_{ec}$ and $m'_{ec}$ \\ 
  $M_i$, $M'_i$                     & transmitted and received codewords \\
  & associated to the message bit $m_i$ \\
  $i$                                & slot time \\
  $s_i$                             & one time password \\
  $f_i$                              & frequency used at slot time $i$ \\
  $F$, ($A$)                                & number of available (jammed) frequencies \\
  $E(f_i)$                          & physical layer function for estimating \\ 
  & the energy of the frequency slot $f_i$ \\
  $\tau$                           & threshold for signal/jamming detection \\
  $H_1(\cdot)$,\  $H_2(\cdot)$ & cryptographically secure hash functions \\
  $p_A$                            & proactive jamming probability \\
  $n$                               & codeword length \\
  $C_n^\Delta$                  & error correction code \\
  $ECC(\cdot)$         & $C_n^\Delta$ encoding algorithm \\
  $ECC^{-1}(\cdot)$         & $C_n^\Delta$ decoding algorithm \\
  $ENC(\cdot)$ & symmetric encryption algorithm \\
  $DEC(\cdot)$ & symmetric  decryption algorithm \\
  $\overline{p}_A$           & maximum jamming probability resiliency \\
  $\epsilon$                   & probability of successful message jamming \\
  $|\cdot|$                      & size of the bit-string \\
\end{tabular}
\end{table}
}
\\
\\
\subsection{Transmitter and Receiver: Software and Hardware Assumptions}
\label{sec:transmitter-receiver}
\begin{figure}
  \centering 
  \includegraphics[height=20mm, width=80mm]{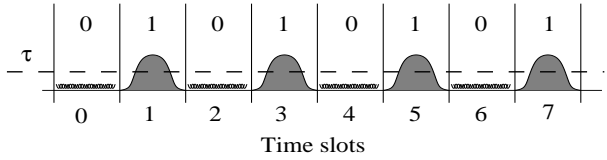}
  \caption{Energy detection capability of the receiver: At each time slot, the receiver
    translates the measured channel energy (RSS) to a bit value, i.e., it decides for
    ``0'' when the measured energy is under the threshold $\tau$, ``1'' otherwise.}
  \label{fig:simple_scheme}
\end{figure}

{\bf Transmitter.} We consider a standard off-the-shelf radio transmitter such as a WiFi,
or a GPRS/UMTS radio device. Without loosing of generality, in the following we consider a
radio technology characterized by $F=124$ different communication channels (like in the
GSM-850).
At each time slot $i$, the transmitter chooses a pseudo-random frequency $f_i
\xleftarrow{\$}[f_0, \ldots, f_{F-1}]$, and as it will be clear in the following, it
decides whether to stay silent or to transmit the message $m$, i.e., $transmit(m, f_i)$.
\\
\\
{\bf Receiver.} In our model, the receiver might reconstruct the transmitted message $m$
in mainly two ways: by either simply receiving (during a time slot) the transmitted
message $m$ or, as it will be clear in the following, by estimating the energy associated
to the frequency $f_i$. In this latter case, recovering $m$ requires multiple time
slots 
--- one of such estimations {\em per} time slot.
\\
We envisage a very simple receiver equipped with a radio and able to estimate the
\emph{received signal strength}, hereafter RSS, i.e., we assume $R$ is provided with a
radio physical layer function $E(f_i)$ which returns an average estimation of the RSS
values experienced during the frequency slot $f_i$. RSS estimation is a common feature in
all the radio devices in order to implement the medium access control. RSS provides an
estimation of the current channel energy. Note that recent papers have leveraged this
information to detect the presence of a jamming signal \cite{capkun_tosn2010}
\cite{trappe}, i.e., when a powerful jamming attack is performed, the receiver experiences
high RSS values.
\\
\\
Therefore, in addition to the standard receiving behavior, our receiver also senses and
logs (into $m'$) whether the energy associated to that frequency exceeds a given threshold
$\tau$.  As a toy example, let us consider Fig.~\ref{fig:simple_scheme}. For each time
slot $i \in [0, \ldots, 7]$, the receiver assesses the channel status and sets $m'_i=1$ if
the RSS overcomes the threshold $\tau$ (this is the case if there is a transmission,
jamming, or environmental noise), while it sets $m'_i=0$ if it senses only noise floor (RSS
under $\tau$).
Communicating leveraging radio silence leads to the following definition:
\begin{definition}
\label{def:silent}
We refer to slot $i$ as a {\bf silent} slot if the energy detected by the receiver on the
associated frequency $f_i$ is below the threshold $\tau$.
\end{definition}

\subsection{Adversarial model}
\label{sec:adversarial-model}
\begin{figure}
  \centering 
  \includegraphics[height=40mm, width=80mm]{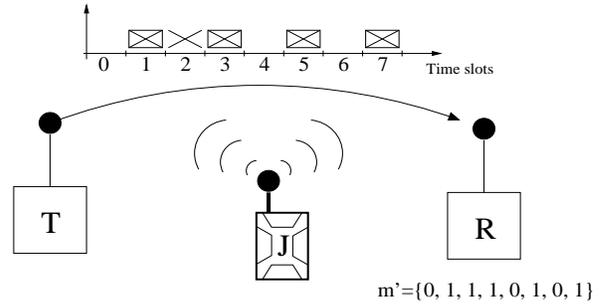}
  \caption{Communication scenario: The transmitter ($T$) delivers a sequence of messages
    to the receiver ($R$). The jammer (\A) successfully jams all the messages. $R$ logs to
    $m'$ the status of each time slot: 1 for jammed, 0 otherwise.}
  \label{fig:model}
\end{figure}

We confront our solution against what, to the best of our knowledge, is the most powerful
adversary presented in the literature. In particular, our adversary (\A), combines the
capability of the proactive jammer, of the reactive jammer, and also network wide
eavesdropping capabilities. In particular:
\begin{itemize}
\item \A\ as a {\bf global eavesdropper}. It is able to eavesdrop all the communications
  in the network. In order to achieve this, \A\ might deploy multiple eavesdropping
  stations all over the network, and moreover, we assume each station is able to monitor
  the overall radio spectrum.
\item \A\ as a {\bf reactive jammer}. It is able to sense the on-going communication and
  to jam it instantaneously, while at the same time switching between the sensing and the
  jamming procedures. Note that considering this powerful (but yet realistic) type of adversary we assume a conservative stance as
  for the security of communications.
  To the best of our knowledge, this is the strongest adversarial configuration
  ever assumed in literature for a reactive jammer.
\item \A\ as a {\bf proactive jammer}. At each time slot, \A\ randomly chooses $A$ among
  the $F$ available frequencies and jams them.
\end{itemize}
Therefore, at each time slot, if $T$ is performing a transmission, \A\ successfully jams
it (whatever the transmission frequency is). Otherwise, \A\ jams $A$ out of the $F$
available frequencies.
\\
\\
As a toy example, let us assume $F=1$ and the communication scenario in
Fig.~\ref{fig:model}. The transmitter sends to the receiver 4 messages and the jammer
successfully jams all of them at time slots $i \in \{1, 3, 5, 7\}$ (reactive jamming).
Moreover, \A\ generates a jamming signal during the time slot $2$ (proactive
jamming). Proactive jamming might appear useless in our adversarial model (all the
communications are already assumed as successfully jammed), nevertheless, as it will be
clear in the following, the silent slots are important to our solution and therefore a
proactive jammer has an incentive in jamming them.
%
\section{{\em \large \bf \sig } preliminaries}
\label{sec:first-look-at}
In this section we introduce the rationales of
the \sig\ protocol, while a detailed description will be provided in the following
section. Let us assume $T$ has to transmit an 8 bits message ($L_m=8$), that is: $m=\{0,
1, 0, 1, 0, 1, 0, 1\}$ (Fig.~\ref{fig:simple_scheme}). The transmitter behaves as follows:
at each time slot $i$ with $0 \le i \le 7$, if $m_i == 1$ than $T$ transmits the whole
$m$, otherwise it waits for the next time slot.
\\
As for the receiver, there can be tree cases: if $R$ correctly receives the message $m$
(benign scenario) than it stops the \sig\ protocol, otherwise, the receiver sets $m'_i=1$
if the RSS exceeds the threshold $\tau$, otherwise $R$ sets $m'_i=0$. However, recalling
Section \ref{sec:adversarial-model}, \A\ (being also a perfect reactive jammer) is assumed
to jam all the slots used by $T$ to transmit the message, as well as (being a proactive
jammer) a few more randomly selected slots (where some of them could be silent slots).
Therefore, let us consider again Fig.~\ref{fig:model}: \A\ successfully jams the slots $i
\in \{1, 3, 5, 7\}$ and also slot $2$ (this latter one was intended by $T$ to be a silent
slot).  We observe that, although all the messages are successfully jammed, after 8 time
slots $R$ is able to recover the bit string $m' = \{0, 1, 1, 1, 0, 1, 0, 1\}$ that differs
from $m$ for only one bit --- the silent slot ($i=2$) jammed by \A.
\\
Similar to the Definition \ref{def:silent}, we define
an {\em active} slot as follows:
\begin{definition}
  \label{def:active}
  Slot $i$ is an {\bf active} slot if, on the associated frequency $f_i$, the transmitter
  $T$ is carrying out an active communication by transmitting a message.
\end{definition}

{\bf Silence is Golden.} The \sig\ protocol interleaves silent slots with active
slots---slot $i$ will be an active one if $m_i==1$, a silent one otherwise.  Both of them
are fundamental for the successful message transmission.  In particular, while active
slots carry the message (or the ``1''s of the message, if the frequency $f_i$ is jammed)
the silent slots carry the ``0''s of the message (or an error if the frequency $f_i$ is
jammed).
\\
\\
In the following, we show how multiple transmissions can be leveraged to mitigate the
(proactive) jamming activity. This feature, combined with channel-idiosyncratic
error-correcting code capabilities enable the full recovery of the original message ($m$).
\subsection{Leveraging frequency-hopping}
\label{sec:enabl-freq-hopp}
\begin{figure}
  \centering 
  \includegraphics[height=50mm, width=80mm]{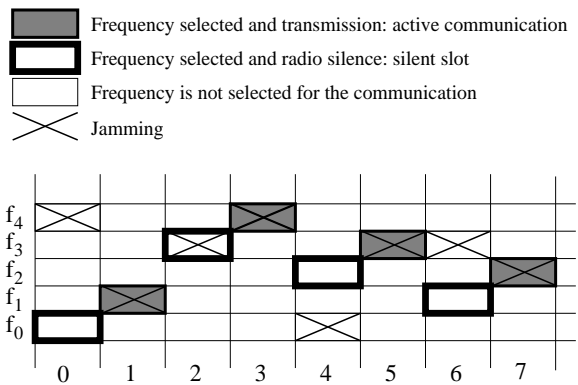}
  \caption{Communication by frequency hopping: $T$ and $R$ share a secret ($s_0$) that is
    leveraged to generate the frequency hopping sequence.}
  \label{fig:freq_hop}
\end{figure}

In the following, we refine the baseline communication scheme above introduced.
\\
As stated before, $T$ and $R$ choose in a pseudo-random fashion the current communication
channel within a set of $F \ge 1$ frequencies. Increasing $F$ makes the proactive jamming
of the current communication channel more difficult, since a larger $F$ decreases the
probability of \A\ to jam the silent slots (communicating the ``0''s of $m$).
\\
In detail, $T$ and $R$ implement a frequency hopping scheme \cite{Popovski} that makes the
current communication frequency unpredictable to the entities that do not share the
initial secret $s_0$. A few solutions have been proposed in order to generate a
pseudo-random (shared) frequency starting from a shared secret. In this work, we adopt the
following formula:
\begin{equation*}
  f_i = H_1(s_i\ |\ i) \mod{F}
\end{equation*}
where $H_1(\cdot)$ is a cryptographically secure hash function, e.g., SHA-1 \cite{sha1},
$i$ is the current time slot, $s_i$ is the shared secret at time slot $i$ and, finally,
$F$ is the total number of available frequencies.
\\
Figure \ref{fig:freq_hop} shows an example of transmission of an 8 bits
message. 
In particular, the frequency hopping sequence is constituted by $\{f_0, f_1, f_3, f_4,
f_2, f_3, f_1, f_2\}$, while the bits involved in the communications are 
$m=\{0, 1, 0, 1, 0, 1, 0, 1\}$.
Each of the above frequency $f_i$ might experience one of the following three different states: silence (white box)---that is,   the sender 
intends to send a 0---; pure jamming (cross)---that is, the frequency is not used by $T$, but it is jammed by \A\---;
or, jammed transmission (grey box with cross). We recall that \A\ is able to jam all the
transmissions that appear in the radio channel, therefore none of the messages sent by the
transmitter will be correctly received by the receiver.
Nevertheless, at each time slot the receiver could still retrieve the bit-message $m_i$ by
assessing the status of the current frequency slot $f_i$.  In fact, when $m_i==1$: $T$
transmits the message, \A\ jams it, and finally, $R$ detects that the RSS associated to
$f_i$ exceeds $\tau$ (indeed, frequency $f_i$ has been jammed) and sets $m'_i==1$.
Whereas, for each $m_i==0$: $T$ selects the radio channel $f_i$ but does not transmit the
message.  $R$ monitors $f_i$ and if no power is detected, decides for $m_i'=0$. This is
the case for time slots $i \in \{0, 4, 6\}$. However, we observe that \A\ might jam a
silent slot (frequency $f_i$ used for a silent slot is randomly selected by the jammer as
well), as it happens in our example for time slot $i=2$ ---causing a one bit error
($m'_2$) in the received sequence $m'$.
Therefore, assuming $F$ available frequencies and an adversary able to jam $A$ of them at
each time slot, the probability for a silent slot to be jammed (flipping a bit from 0 to
1) is given by $p_A = \frac{A}{F}$. Moreover, we highlight that, regardless of \A\
activities, the receiver is anyway able to retrieve at least all the ``1''s of the
message. 
%
\subsection{Binary asymmetric error correcting codes}
\label{sec:binary-asymm-error}
\begin{figure}
  \centering 
  \includegraphics[height=30mm, width=32mm]{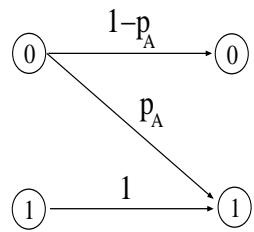}
  \caption{The inverted Z-channel: $P(0 \rightarrow 1) = p_A$ and $P(1 \rightarrow 1) =
    0$}
  \label{fig:zch}
\end{figure}

As stated in Section \ref{sec:enabl-freq-hopp}, \A\ can always prevent the correct
reception of the messages, given its perfect reactive jamming capabilities. However, $R$
is still able to recover all the ``1''s of $m$.
We stress that \A\ cannot prevent the (active) communication of the ``1''s, in fact the
only way to achieve this is to remove the message from the radio spectrum, e.g.,
generating an inverse waveform to have the RSS sensed by $T$ resulting below the threshold
$\tau$. However, in the literature this feature is considered very difficult to achieve
\cite{baird}; therefore, in the following we will assume such an event as
impossible---i.e. experiencing a bit transition from ``1'' to ``0'' has associated
probability 0.  Nevertheless, \A\ has a probability $p_A$ to jam a silent channel, that is
changing the bit value from ``0'' to ``1''.
\\
Communication channels characterized by an asymmetric probability to experience a
transition between zeros and ones, such as the one above described, are said \emph{binary
  asymmetric channels} \cite{klove}, hereafter BAC. In particular, BAC characterized by
$P(0 \rightarrow 1) = p_A$ and $P(1 \rightarrow 1) = 0$ are said \emph{inverted
  Z-channels} \cite{klove}, see Fig.~\ref{fig:zch}.
The inverted Z-channel shows a perfect fit 
to model our communication channel.  Mostly important, the error correcting codes (ECC)
specifically designed for this channel might be used to recover the error bits due to
jamming over silent slots.
\\
Let $x$ and $y$ be two bit strings (codeword) of $n$ bits each belonging to the code
$C$. Let $\delta$ be the asymmetric distance, i.e., the number of $i's$ such that $m_i=0$
and $m'_i=1$.  Let also $\Delta = \min_{\{x,y \in C, x \neq y\}} \delta(x,y)$ be the
minimum asymmetric distance.  A fundamental theorem of the ECC theory \cite{delsarte}
follows:
\par\vspace{-0.5cm}
\begin{thm}
  An asymmetric binary code of minimum asymmetric distance $\Delta$ is capable of
  correcting $t$ or fewer errors of type $0 \rightarrow 1$, where t is fixed and satisfies
  $t \le \Delta - 1$.
\end{thm}
\par\vspace{-0.5cm}
Authors in \cite{delsarte} provide several constructions for asymmetric binary error
correction codes $C_n^\Delta$, given the codeword length $n$ and the minimum asymmetric
distance $\Delta$.
\\
In the following, we consider the most resilient configuration, i.e., $n = \Delta$, that
is able to correct up to $n-1$ errors. Now, let us assume a bit string $m_e$ of $L_m$
bits. In order to be resilient to $n-1$ consecutive jamming hits, each bit $m_{e_i}$ of
the message $m_e$ is encoded with a codeword $M_i$ of length $n$ bits, i.e., repeated $n$
times.
We adopt the following notation: $m_{ec} = ECC(m_e)$ where $m_e$ and $m_{ec}$ are
bit-strings of $L_m$ and $L_c=nL_m$ bits, respectively, and $m_{ec}= \{M_0, \ldots,
M_{L_m-1}\}$. In particular, the $i^{th}$ bit of the message $m_e$ is encoded into the $n$
bits codeword $M_i$, i.e., $M_i = \{0, \ldots, 0\}$ if $m_{e_i} == 0$, else $M_i = \{1,
\ldots, 1\}$ if $m_{e_i} == 1$.
Conversely, the receiver runs the decoding algorithm, i.e., $m'_e = ECC^{-1}(m'_{ec})$,
where $m'_{ec}= \{M'_0, \ldots, M'_{L_m-1}\}$ and decides for $m_{e_i}==1$ if $M'_i = \{1,
\ldots, 1\}$, otherwise it sets $m'_{e_i}==0$.
%
%
\section{The {\em \large \bf \sig} protocol}
\label{sec:protocol}
\begin{figure}
  \centering 
  \includegraphics[height=30mm, width=85mm]{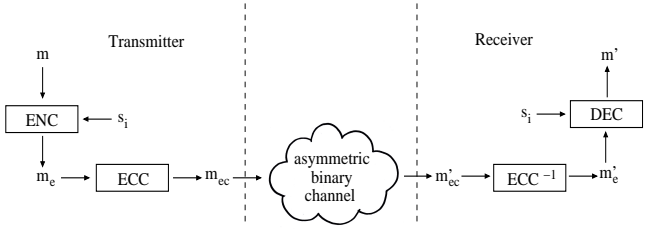}
  \caption{The \sig\ protocol: Transmitter and Receiver model. $|m| = |m_e| = |m'| =
    |m'_e| = L_m$ and $|m_{ec}| = |m'_{ec}| = L_c$. }
  \label{fig:sig}
\end{figure}

The \sig\ protocol guarantees both confidentiality and integrity of the transmitted
message. Each message $m$, before being transmitted, is encrypted with a one-time-password
(OTP) $s_i$, i.e., $m_e = ENC(m, s_i)$, where $ENC(\cdot)$ is a symmetric encryption
algorithm (such as AES\cite{aes}), encrypting message $m$ with key $s_i$.  At each time
slot a new (shared) secret key $s_i$ is generated by both parties computing $s_i =
H_2(s_{i-1})$, with $i \ge 1$ --- where $H_2(\cdot)$ is a cryptographically secure hash
function \cite{sha1}. Conversely, the receiver computes $m'= DEC(m'_e, s_i)$, where
$DEC(\cdot)$ is the symmetric decryption
algorithm. 
\\
Figure \ref{fig:sig} shows the overall model, while details are provided in Section
\ref{sec:transmitter} and \ref{sec:receiver}, respectively.
\subsection{Transmitter}
\label{sec:transmitter}
Algorithm \ref{sec:send} shows the sequence of steps performed by $T$ in order to transmit
the bit-string $m$ to the receiver $R$. The transmitter algorithm needs as input the
message $m$ and the shared (with $R$) secret $s_{i-1}$ ($i \ge 1$). Assuming the current
time slot as $i$, the first step is to generate a new OTP, i.e., $s_i$ at line 9. The new
shared secret $s_i$ will be used to both encrypt the message $m$ and generate the next
transmission frequency. The message encryption is obtained by means of $m_e = ENC (m,
s_i)$ while the encoding is performed as: $m_{ec} = ECC(m_e)$, obtaining a bit-string of
$L_c = n L_m$ bits (line 11). Now, for each time slot $i$ (in the next $L_c$ slots), the
transmitter decides whether to transmit the message $m_e$ (line 16) or to stay silent
(according to the bit value $m_{ec}[i]$ (line 18--19)). Each frequency slot is chosen
according to $f_i = H_1(s_i\ |\ i) \mod{F}$ (line 13).
\\
Eventually, $m$ is transmitted to $R$ after $n L_m$ time slots with an average of
$\frac{nL_m}{2}$ transmissions, assuming the distribution of the bit values $\{0,1\}$ as
uniform\footnote{We recall that $m_e$ is encrypted, therefore we
  would expect a uniform bit distribution \cite{Soto}.}.

%
\begin{algorithm}[t]
  \SetKwInOut{Input}{Input}
  \caption{Transmitter side}
  \label{sec:send}
  \Input{Shared secret: $s_{i-1}$, Message: $m$}
  \BlankLine
  {\bf let } $m$ be a bit-string of $L_m$ bits. \\
  {\bf let } $s_{i-1}$ be the shared secret with $R$ at time slot $i-1$. \\
  {\bf let } $i$ be the current time slot. \\
  {\bf let } $m_e$ be a bit-string of $L_m$ bits. \\
  {\bf let } $m_{ec}$ be a bit-string of $L_c$ bits. \\
  {\bf let } $ECC(\cdot)$ be the ECC encoding algorithm. \\
  {\bf let } $ENC(\cdot)$ be a symmetric encryption algorithm. \\
  {\bf let } $F$ be the number of available frequencies.\\
  \BlankLine
  \tcc{\footnotesize{Generate a new OTP.}}
  $s_i = H (s_{i-1})$\;
  \tcc{\footnotesize{Encrypt the message $m$ into $m_e$}}
  $m_e = ENC (m, s_i)$\;
  \tcc{\footnotesize{Encode the bit-string $m_{e}$ into the bit-string $m_{ec}$}}
  $m_{ec} = ECC(m_e)$\;
  \For{$i = 1 \ldots L_c$}{
    \tcc{\footnotesize{Select the communication frequency}}
    $f_i = H_1(s_i\ |\ i) \mod{F}$\;
    \tcc{\footnotesize{Retrieve one bit from $m_{ec}$}}
    $c = m_{ec}[i]$\; 
    \If{$c == 1$}
    {
      \tcc{\footnotesize{Transmit $m_e$ at frequency $f$}}
      transmit($m_e$, $f_i$)\;
    }
    \Else
    {
      \tcc{\footnotesize{Wait till the next time slot}}
    }
  }
\end{algorithm}
\subsection{Receiver}
\label{sec:receiver}
The receiver algorithm (Algorithm \ref{sec:receive}) starts by synching with $T$ on the
new shared secret key, i.e., $s_i = H(s_{i-1})$, where $i$ is the current time slot (line
10). If all the transmitted messages have been successfully jammed by \A, for each of the
next $L_c$ time slots, $R$ performs three steps: frequency selection (line 13), channel
energy measurement (line 20), and decision on the value of the current bit (line 21---27).
\\
The receiver syncs with the transmitter on the correct frequency by means of $f_i =
H_1(s_i\ |\ i) \mod{F}$. The receiver performs the message reception by means of $m'_e =
receive(f_i)$, decrypts $m'_e$ obtaining $m'$, and finally, if the integrity check of $m'$
is successful ($m' \in \Phi$), it sets the $rx$ variable to true (line 17). Nevertheless,
our adversarial model assumes that none of the message can be received correctly, i.e.,
\A\ is able to jam all the active communications.
\\
Therefore, $R$ leverages the channel energy in order to reconstruct the transmitted
message. The receiver retrieves the estimation of the energy on the current frequency slot
$f_i$ by means of the radio function $e = E(f_i)$. If the estimated energy $e$ overcomes
the threshold $\tau$, the receiver sets $m'_{ec}[i] = 1$, otherwise $m'_{ec}[i] =
0$. Eventually, after $L_c$ time slots, the receiver firstly decodes the collected bits
($m'_{ec}$) into the bit-string $m'_e$, and subsequently decrypts $m'_e$ obtaining the
message $m'$.
\\
Finally, the receiver checks for the message integrity, i.e., $m \in \Phi$, and returns
$m=m'$ if the message is correct, otherwise error (lines 33---38).
\\
\\
{\bf Packet loss.} Generally speaking, packet loss is due to radio noise that corrupts the
transmitted packets \cite{rappaport}.  In our communication model the correct reception of
a bit at slot $i$ depends on the energy sensed over frequency $f_i$.  Hence, in principle
it could be possible that random energy fluctuations in the channel might produce
destructive interference, causing the energy associated to frequency $f_i$ to be below the
threshold $\tau$, eventually generating a crossover $1 \rightarrow
0$. 
However, in our reference scenario (that is, assuming the presence of \A), we observe that
our adversarial model involves a reactive jammer that (when a transmission is sensed) jams
the overall network with a very powerful signal. Since the energy over frequency $f_i$ is
increased by the jammer, the jammer itself makes the probability of a crossover $1
\rightarrow 0$ negligible.
\\
Finally, we stress that in any case a corrupted message cannot be accepted as a genuine
one. Indeed, the receiver $R$ eventually checks for the integrity of the message (line 33
in Algorithm \ref{sec:receive}), and discards the message if it does not pass the check.
\begin{algorithm}
  \SetKwInOut{Output}{Output}
  \SetKwInOut{Input}{Input}
  \caption{Receiver side}
  \label{sec:receive}
  \Input{Shared secret: $s_{i-1}$} 
  \Output{Message $m'$ if $m' = m$, otherwise {\bf error}} 
  \BlankLine
  {\bf let } $m'$ be a bit-string of $L_m$ bits. \\
  {\bf let } $s_{i-1}$ be the shared secret with $R$ at time slot $i-1$. \\
  {\bf let } $i$ be the current time slot. \\
  {\bf let } $m'_e$ be a bit-string of $L_m$ bits. \\
  {\bf let } $m'_{ec}$ be a bit-string of $L_c$ bits. \\
  {\bf let } $ECC^{-1}(\cdot)$ be the ECC decoding algorithm. \\
  {\bf let } $DEC(\cdot)$ be a symmetric decryption algorithm. \\
  {\bf let } $F$ be the number of available frequencies.\\
  {\bf let } $\tau$ be the energy decision threshold.\\
  \BlankLine
  \tcc{\footnotesize{Generate a new OTP.}}
  $s_i = H (s_{i-1})$\;
  $i = 0$;\  $rx={\bf false}$\; 
  \While{$i < L_c$ {\bf and} not(rx)}{
    \tcc{\footnotesize{Select the communication frequency}}
    $f_i = H_1(s_i\ |\ i) \mod{F}$\;

    \tcc{\footnotesize{Receive the message $m'_e$ at the freq. $f_i$}}
    $m'_e$=receive$(f_i)$\;

    \tcc{\footnotesize{Decrypt $m'_e$ with $s_i$}}
    $m' = DEC (m'_e,\ s_i)$\;

    \tcc{\footnotesize{Check $m'$ integrity}}
    \If{$m' \in \Phi$}
    {
       \tcc{\footnotesize{Message has been correctly received.}}
      $rx=$ {\bf true}\; 
    }
    \Else
    {
      \tcc{\footnotesize{Retrieve the channel energy.}}
      $e = E(f_i)$\;
      \If{$e \ge \tau$}
      {
        $m'_{ec}[i] = 1$\;
      }
      \Else
      {
        $m'_{ec}[i] = 0$\;
      }
    }
  }
  \If{not(rx)}
  {
    \tcc{\footnotesize{Decode $m'_{ec}$ into $m'_e$}}
    $m'_{e} = ECC^{-1}(m'_{ec})$\;    
    \tcc{\footnotesize{Decrypt $m'_e$ with $s_i$}}
    $m' = DEC (m'_e, s_i)$\;
  }
  \tcc{\footnotesize{Check $m'$ integrity}}
  \If{$m' \in \Phi$\ {\bf or} rx}
  {
    {\bf return} $m'$\;
  }
  \Else
  {
    {\bf return} {\bf error}\;
  }
\end{algorithm}
\subsection{Wrap up}
\label{sec:putt-everyth-togeth}
\begin{figure*}
  \centering 
  \includegraphics[height=70mm, width=170mm]{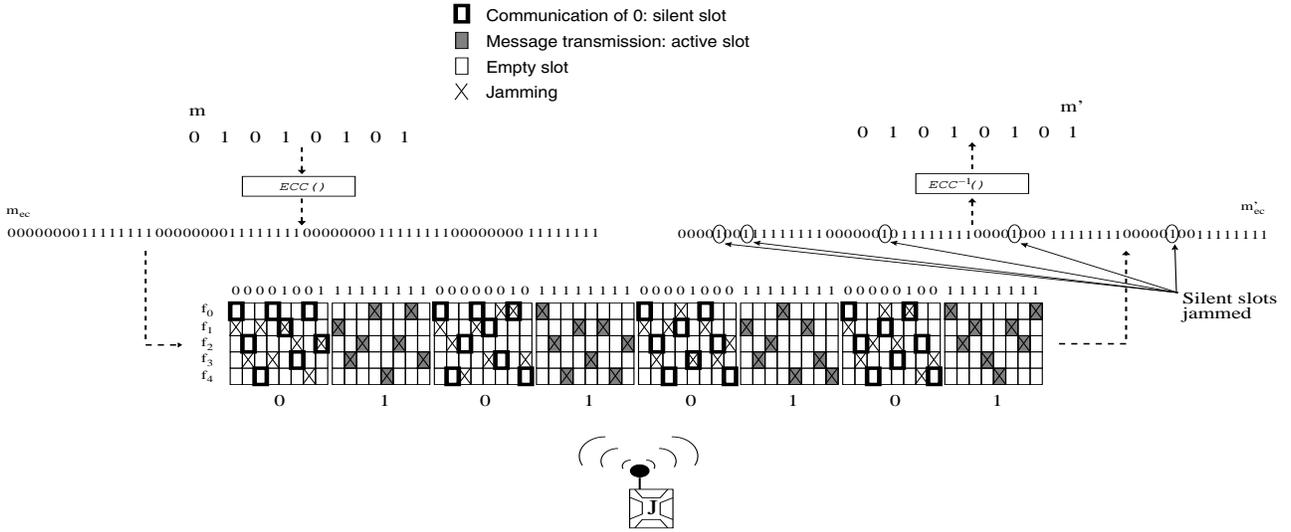}
  \caption{An example of message transmission, jamming, and reception: the bit-string $m = \{1, 0,
    1, 0, 1, 0, 1, 0\}$ is encoded into the bit-string $m_{ec}$, and subsequently
    transmitted into a radio spectrum with $F=5$ frequencies. The jammer jams all the
    messages and a few silence slots flipping one or more bits (from 0 to 1). Finally, the
    receiver recovers the original bit-string $m$ by leveraging the error correcting
    code.}
  \label{fig:toy_ex}
\end{figure*}

The \sig\ protocol combines two key elements that make itself robust to jamming: (i)
frequency hopping makes the communication of the ``0''s unpredictable, while (ii) the
(active) communication of ``1''s cannot be prevented by \A\ (i.e. the transmission of
``1''s is transparent to jamming).
\\
We consider a simple example of the \sig\ protocol in Fig.~\ref{fig:toy_ex}. In order to
ease the discussion, we do not consider the encryption step, therefore the bit-string $m$
is directly encoded into the bit-stream $m_{ec}$. Further, we assume a code $C^\Delta_n$,
such that $\Delta=n=8$, and consequently able to recover $t \le \Delta-1=7$ errors, see
Section \ref{sec:binary-asymm-error} for the details. Therefore, the initial bit-string
$m=\{0, 1, 0, 1, 0, 1, 0, 1\}$ of length $L_m=8$ bits is encoded into the bit-stream
$m_{ec}$ of length $L_c = nL_m = 64$ bits. The communication channel is constituted by
$F=5$ frequencies: at each time slot the transmitter and the receiver sync on one of them,
and subsequently, either a transmission or a radio silence is performed as function of the
current value of the bit to be transmitted.
\\
The jammer jams both all the transmitted messages and (proactively) $A=1$ of the $F=5$
frequencies when no communications appear on the radio spectrum. We observe that \A\ hits
the $5^{th}$, $8^{th}$, $23^{rd}$, $37^{th}$, and the $54^{th}$ silent slot, and
consequently, the bit-string $m'_{ec}$ differs from $m_{ec}$ of 5 bits. Nevertheless, the
ECC code is able to correct up to $t=7$ errors per codeword, and eventually the received
message $m'_{ec}$ allows to recover $m$.
\\
\\
We stress that under a standard proactive adversary (that is, an adversary that with not
zero probability fails to jam an active communication), the \sig\ protocol delivers the
message $m$ with the first not-jammed active communication. Whereas, our adversary \A\
successfully jams all the active communications, and therefore \sig\ accomplishes the
correct delivery of a single bit (``1'') per active communication. Although \A\ cannot
prevent the delivery of the ``1''s, it can jam the transmission of the ``0''s, changing
their value to ``1''.
\section{Performance evaluation}
\label{sec:perf-eval}
In the following, we present the performance analysis of the \sig\ protocol. We start our
analysis from a theoretical point of view providing a closed formula for the probability
that $R$, having derived $m'$ from the received message $m'_{ec}$, correctly recovers the
message $m$ originally sent by $T$ (we will refer to this probability as $P(m'=m)$). Such
a probability will be dependant on the (proactive) jamming probability $p_a$. We will
assume, coherently with our adversary model, that the reactive capabilities of the jammer
allow it to jam all the transmissions of $T$---hence, all the 1s of message $m$ are
correctly received (cfr. Section \ref{sec:adversarial-model}).  Subsequently, we show and
discuss the results of an extensive simulation campaign---confirming our theoretical
findings and the quality and viability of our proposal.
\subsection{Theoretical analysis}
\label{sec:theoretical-analysis}
\begin{figure}
  \centering 
  \includegraphics[height=30mm, width=70mm]{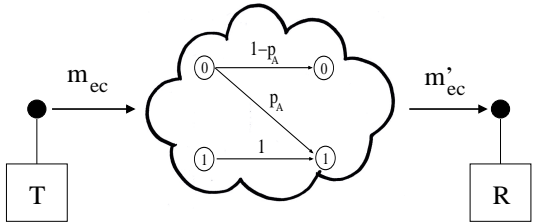}
  \caption{The communication reference model for the theoretical analysis.}
  \label{fig:theo_model}
\end{figure}
Figure \ref{fig:theo_model} recaps our communication reference model for the theoretical
analysis. In particular, we recall that each bit of the message $m$, i.e. $m_i$, is
encoded into a codeword $M_i$ of $n$ bits (as {\em per} Section
\ref{sec:binary-asymm-error}), that is, $m_{ec} = ECC(m)$, where $m_{ec} = \{M_0, \ldots,
M_{L_m-1}\}$. The probability $P(M_i=M'_i)$ that the codeword $M_i$, with $i \in [0,
L_m-1]$, is correctly delivered to $R$ assuming the channel model of
Fig.~\ref{fig:theo_model}, is:
\begin{equation}
  P(M_i=M'_i) = P\Big(\delta(M_i, M'_i) < n\Big) 
  \label{eq:M_Mf}
\end{equation}
where $\delta(M_i, M'_i)$ is the asymmetric distance computed between $M_i$ and $M'_i$,
i.e., the number of ``0''s belonging to $M_i$ that change their value to ``1'' in the
bit-string $M'_i$. We recall that, according to our communication model---justified in
previous sections and synthesized in Fig. \ref{fig:theo_model}---, the only possible bit
crossover is $0 \rightarrow 1$, while $1 \rightarrow 0$ is not possible.  Therefore, since
the frequencies jammed by a proactive jammer in any time slot are independent from the
frequencies jammed in other time slots, Eq.~(\ref{eq:M_Mf}) can be rewritten as:
\begin{equation}
  P(M_i=M'_i)  =  1 - P\Big(\delta(M_i, M'_i) = n\Big) 
  \nonumber 
\end{equation}
The probability to experience exactly one crossover $0 \rightarrow 1$ at a given time slot
of codeword $M_i$ is given by the probability $p_a$ that \A\ successfully jams exactly
that time slot (out of the $n$ silent slots) belonging to the codeword $M_i$.
Therefore, the probability to have the codeword $M_i$ correctly decoded at the receiver, yields:
\begin{equation}
  P(M_i=M'_i)  =  1 - p_a^n   \label{eq:M_Mf_2} 
  \nonumber 
\end{equation}
Further, the probability to correctly deliver the bit-string $m_{ec}$, i.e., $P(m_{ec} =
m'_{ec})$, can be computed as:
\begin{equation}
  P(m_{ec}=m'_{ec})  =   \prod_{i=0}^{L_m-1} P(M_i = M'_i) 
  \nonumber
\end{equation}
Assuming the message $m$ enjoys a uniform distribution of zeros and ones (similar
considerations expressed in footnote 1 do support this assumption) and recalling that, as
justified in Section \ref{sec:receiver} $P(M_i = M'_i \ |\ m_i = 1) = 1$ ---no crossovers
$1 \rightarrow 0$ occur---, the above equation can be rewritten as:
\begin{eqnarray}
  P(m_{ec}=m'_{ec})  & = &  \prod_{i=0}^{\frac{L_m-1}{2}} P(M_i = M'_i \ |\ m_i = 0) \nonumber \\ 
  & = & (1 - p_a^n)^\frac{L_m}{2}   \label{eq:m_mf_pa}
\end{eqnarray}
The probability that $R$ could recover an $L_m$ bits long message $m$ sent by $T$ and
encoded with an ECC code $C_n^\Delta$, yields:
\begin{equation}
  P(m=m') = \left(1 - p_a^n \right)^\frac{L_m}{2} \ge e^{ -p_a^n L_m}
  \label{eq:m_mf}
\end{equation}
Therefore, if we set to $\epsilon$ the upper bound on the probability for \A\ to
successfully jam the message $m$ (i.e. $P(m' \neq m) \le \epsilon$), the maximum jamming
probability $\overline{p}_a$ the protocol is resilient to can be computed as:
\begin{equation}
  \overline{p}_a = \sqrt[n]{-\frac{1}{L_m}\ln(1 - \epsilon)}
  \label{eq:A_F_bound}
\end{equation}
\subsection{Simulation results}
\label{sec:simulation-results}
\begin{figure}
  \centering 
  \includegraphics[height=60mm, width=80mm]{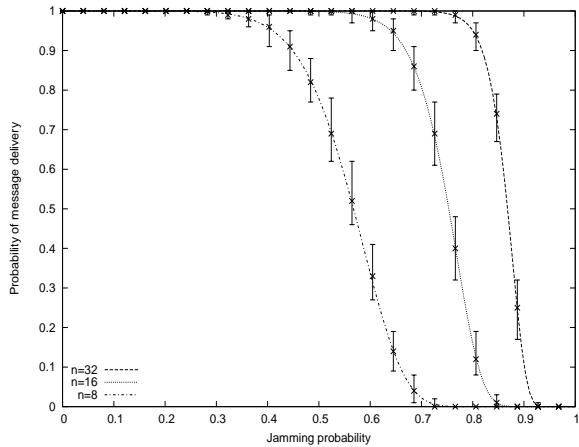}
  \caption{Message delivery probability ($P(m=m')$) with the \sig\ protocol: the message
    length is $L_m=128$ bits, while the jamming probability has been obtained by fixing
    $F=124$ and varying $A$ in $[0, \ldots, 124]$. We consider both experimental
    (errorbars) and theoretical (curves) results for different codeword lengths, i.e., $n
    \in \{8, 16, 32\}$.}
  \label{fig:SiG_A_100_N_100_100_100}
\end{figure}

\begin{figure}
  \centering 
  \includegraphics[height=60mm, width=80mm]{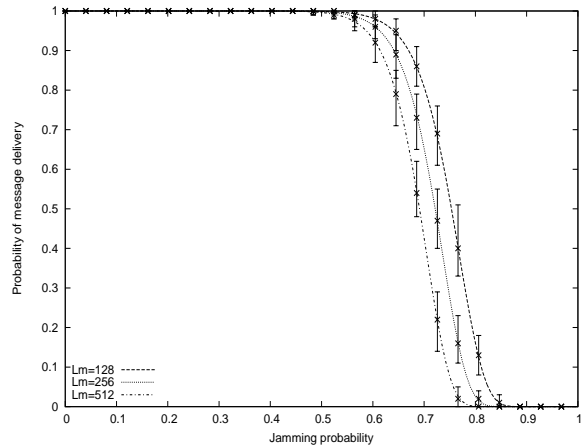}
  \caption{Message delivery probability ($P(m=m')$) with the \sig\ protocol: the message
    length spans in the range $L_m \in \{128, 256, 512\}$ bits, while the jamming
    probability has been obtained by fixing $F=124$ and varying $A$ in $[0, \ldots, 124]$. We consider both
    experimental (errorbars) and theoretical (curves) results for a fixed codeword length
    $n = 16$.}
  \label{fig:SiG_A_100_10_Lm_100_100}
\end{figure}

\begin{figure}
  \centering 
  \includegraphics[height=60mm, width=80mm]{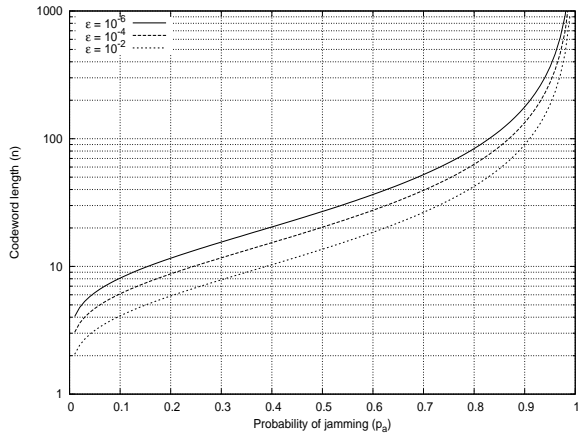}
  \caption{Choosing the codeword length ($n$) as function of the probability of jamming
    ($p_a$). We fixed the message length $L_m = 128$ bits and considered different values
    for the threshold $\epsilon \in \{10^{-2}, 10^{-4}, 10^{-6}\}$.}
  \label{fig:SiG_theo_n_pa}
\end{figure}

We consider the reference scenario of Fig. \ref{fig:theo_model}, and the transmission of a
message $m$ of length $L_m=128$ bits.  Figure \ref{fig:SiG_A_100_N_100_100_100} shows both
theoretical and simulated results of the \sig\ protocol.  
Errorbars show the quantile 5,
50, and 95 of $10,000$ simulated transmissions of the message $m$. For each configuration,
we derived the jamming probability $p_a$ by setting the number of available frequencies to
the constant $F=124$, while we varied the number of jammed frequencies $A$ from 0 to
124. Moreover, we consider three different codeword lengths, i.e., $n \in \{8, 16,
32\}$. Finally, the pointed curves represent the theoretical predictions provided by
Eq.~(\ref{eq:m_mf}).
\\
Table \ref{tab:bounds} shows the bounds on $\overline{p}_a$ fixing $\epsilon = 0.99$ and
varying $n \in \{8, 16, 32\}$; recalling Eq.~(\ref{eq:A_F_bound}), we can observe in
Figure \ref{fig:SiG_A_100_N_100_100_100} how the bounds perfectly fit the experimental
results. 
\begin{table}
  \caption{Bounds for the maximum jamming probability resiliency ($\overline{p}_a$) varying the
    codeword length $n \in \{8, 16, 32\}$ and fixing $\epsilon= 10^{-2}$. \label{tab:bounds}}
  \centering
\begin{tabular}{ c c }
  $n$ & $\overline{p}_a$ \\
  \hline
  8  & 0.30 \\
  16 & 0.55 \\
  32 & 0.74 \\
\end{tabular}
\end{table}
For instance, note that \sig\ is able to deliver a 128 bit-string with probability at
least $99\%$ ($P(m=m') \ge 1-\epsilon$), using a codeword length of $n=16$ in the presence
of a jammer \A\ which proactively jams the $55\%$ of the available frequencies.  Finally,
we want to stress that our results are obtained assuming that the jamming is performed
successfully on all the active communications and on a subset ($p_a = \frac{A}{F}$) of the
silent radio channels.
\\
\\
{\bf Varying the message length $L_m$.} Figure \ref{fig:SiG_A_100_10_Lm_100_100} shows
both theoretical and simulated results of the \sig\ protocol varying the message length
$L_m \in \{128, 256, 512\}$. 
We fixed the codeword length to $n=16$, and we set the jamming probability by fixing the
number of available frequencies to $F=124$, while we varied the number of jammed
frequencies $A$ from 0 to 124.  Errorbars show the quantile 5, 50, and 95 of $10,000$
simulated transmissions of the message $m$, while the curves are obtained by plotting
Eq.~(\ref{eq:m_mf}).  Recalling Eq.~(\ref{eq:A_F_bound}), we observe that the bounds in
order to guarantee a message delivery with at least $99\%$ probability ($P(m=m') \ge
1-\epsilon)$, are given by $p_a \le \overline{p}_a = \{0.55, 0.53, 0.51\}$ for a message
of $L_m = \{128, 256, 512\}$ bits, respectively, and a codeword length $n=16$.
\\
\\
{\bf Varying the threshold $\epsilon$.} Finally, we consider how the successful message
jamming probability $\epsilon = P(m \neq m')$ affects the performance of the \sig\
protocol. Equation (\ref{eq:A_F_bound}) can be rewritten as function of the codeword
length, yielding:
\begin{equation}
  n = \frac{\ln{(-\frac{1}{L_m} \ln{(1-\epsilon)}})}{\ln{(p_a)}}
  \label{eq:n}
\end{equation}
Figure \ref{fig:SiG_theo_n_pa} shows Eq.~(\ref{eq:n}) varying $p_a$ for different values
of the threshold $\epsilon \in \{10^{-2}, 10^{-4}, 10^{-6}\}$, with a message length
$L_m=128$ bits. For instance, assuming a (proactive) jamming probability
$p_a=\frac{A}{F}=0.8$, we observe that a codeword length $n$ ranging in the interval $[40,
80]$ assures a message delivery probability of $1-\epsilon$, where $\epsilon$ ranges in $[10^{-2},
10^{-6}]$.
\\
\section{Comparison with other \\solutions}
\label{sec:comp-with-other}
In this section we compare our solution with other anti-jamming techniques: Table
\ref{tab:comparison} compares \sig\ with other recent works as function of the adversarial
behavior. Firstly, we recall that ---to the best of our knowledge--- the adversary
considered in this work (\A) is the most powerful ever considered in the literature (see
Section \ref{sec:adversarial-model}). Standard techniques as
\cite{lazoswisec}\cite{capkun_jsac} assume a ``pure'' proactive adversary, and therefore
cannot deal with \A, in fact, both TDBS \cite{lazoswisec} and UFH-UDSSS \cite{capkun_jsac}
are useless against a reactive jammer that promptly interferes with the transmitted
message. In particular, TDBS changes the transmission frequency of the peers according to
a pre-loaded sequence; nevertheless, the simple frequency hopping is useless against \A,
which reactively jams the transmitted message as soon as it appears on the
channel. Similarly, UFH-UDSSS combines both uncoordinated frequency hopping and
uncoordinated direct spread spectrum: although this approach does not need a pre-shared
secret between the peers, a reactive adversary like \A\ disrupts the communications and
prevents  message delivery.
\\
A few solutions have been proposed in order to mitigate the effects of a reactive
adversary. The solution presented in \cite{capkun_tosn2010} introduces a novel technique
to detect a reactive jammer and raise a jamming suspicion alarm. Authors leverage the
combination of bit errors and RSS readings in order to infer on the current presence of a
jamming signal. Although, this solution is optimal for the protection of a reactive alarm
system, it does not solve the problem of communicating in the presence of a reactive
jammer. A similar solution is proposed by \cite{xuan}: nodes that trigger the reactive
jammer are switched off and the messages are routed in order to avoid the nodes close to
jammers. This solution involves mainly the identification of jammers' position and does
not deal directly with the jamming attack, yet authors assume the jammed area is a subset
of the network deployment, and therefore, the proposed solution is not effective against a
\A\ that can jam the whole network. An interesting solution that
directly deals with a reactive jammer is BitTrickle \cite{liu}. The solution leverages the 
delay 
experienced by a jammer 
to switch between the sensing and the jamming phase in order to correctly
deliver a few bits per packet. Although the authors assume a reactive jammer with
unlimited spectrum coverage and transmission power, the proposed solution is not resilient
to \A.  Indeed, in our adversarial model \A\ experiences a theoretically zero delay to
switch between sensing and jamming.
e the frequency slotsused for the transmissions). 
A MAC level solution is proposed in \cite{richa}: authors
design a protocol that guarantees fair channel access probabilities among nodes in the
presence of a reactive jammer. Nevertheless, as for the previous solutions, even AntiJam
\cite{richa} cannot deal with \A, in fact, to the best of our knowledge, none of the
solution proposed in the literature can deal with the reactive jammer introduced in this paper.
%
\\
Finally, the solution presented in \cite{blass} is the only one that can be adopted in
order to deal with a combined proactive-reactive adversary. The solution is mainly based
on a novel mechanical beam-forming design with a fast auto-configuration algorithm, i.e.,
the geometry of a two-element antenna is controlled by an algorithm in order to obtain a
destructive interference for the received jamming signal. Nevertheless, such an approach
cannot deal with multiple deployed adversaries or even against a single mobile adversary:
indeed, antenna cancellation is achieved with respect to only a specific (static)
adversarial position.
\begin{table}
  \caption{Comparison with other solutions as function of the \A's behavior. \label{tab:comparison}}
  \centering
  \begin{tabular}{| l || c | c | c |}
    \hline
    {\bf Name} & \parbox[t]{0.5in}{Proactive \par adversary} & \parbox[t]{0.5in}{Reactive \par adversary} & \parbox[t]{0.6in}{Multiple \par adversaries} \\ 
    \hline
    TDBS \cite{lazoswisec} & $\checkmark$ & $\times$  & $\checkmark$ \\ 
    \hline
    UFH-UDSSS \cite{capkun_jsac} & $\checkmark$ & $\times$ & $\checkmark$ \\
    \hline
    \cite{capkun_tosn2010} & $\times$  & $\checkmark$ & $\checkmark$ \\
    \hline
    \cite{xuan} & $\times$ & $\checkmark$ & $\checkmark$ \\
    \hline
    BitTrickle \cite{liu} & $\times$ & $\checkmark$ & $\times$ \\
    \hline
    AntiJam \cite{richa} & $\times$ & $\checkmark$ & $\checkmark$ \\
    \hline
    \cite{blass} & $\checkmark$ & $\checkmark$ & $\times$ \\
    \specialrule{.2em}{.0em}{.0em}
    $SiG$ & $\checkmark$ & $\checkmark$ & $\checkmark$ \\
    \hline
  \end{tabular}
\end{table}
\section{Conclusions}
\label{sec:conclusions}
In this paper we have introduced a powerful (yet realistic) jammer that is able to reduce
to zero the communication bandwidth between two communicating parties, even when state of
the art anti-jamming solutions are adopted.  To cope with this novel adversary model, we
have introduced a brand new communication protocol: Silence is Golden (\sig).
\\
Implementing a tunable, asymmetric communication channel between communicating parties,
\sig\ is able to restore an effective bandwidth between them.  We have provided a thorough
analysis of the SiG protocol, as well as the results of an extensive simulation campaign
that do support our theoretical findings and the viability of the SiG protocol.
\newpage
\balance
\bibliographystyle{abbrv}
\bibliography{ccs2013}  

\begin{thebibliography}{10}

\bibitem{awerbuch}
B.~Awerbuch, A.~Richa, and C.~Scheideler.
\newblock A jamming-resistant mac protocol for single-hop wireless networks.
\newblock PODC '08, pages 45--54, New York, NY, USA, 2008.

\bibitem{baird}
L.~Baird, W.~Bahn, M.~Collins, M.~Carlisle, and S.~Butler.
\newblock Keyless jam resistance.
\newblock In {\em Information Assurance and Security Workshop, 2007. IAW '07.
  IEEE SMC}, pages 143--150, June.

\bibitem{bhadra}
S.~Bhadra, S.~Bodas, S.~Shakkottai, and S.~Vishwanath.
\newblock Communication through jamming over a slotted aloha channel.
\newblock {\em Information Theory, IEEE Transactions on}, 54(11):5257--5262,
  Nov.

\bibitem{dehghan}
M.~Dehghan, D.~Goeckel, M.~Ghaderi, and Z.~Ding.
\newblock Energy efficiency of cooperative jamming strategies in secure
  wireless networks.
\newblock {\em Wireless Communications, IEEE Transactions on}, (99):1--5, 2012.

\bibitem{delsarte}
P.~Delsarte and P.~Piret.
\newblock Bounds and constructions for binary asymmetric error-correcting codes
  (corresp.).
\newblock {\em Information Theory, IEEE Transactions on}, 27(1):125--128, Jan.

\bibitem{networkCR13}
R.~{Di Pietro} and G.~Oligeri.
\newblock Jamming mitigation in cognitive radio networks.
\newblock {\em IEEE Network Magazine, Special Issue on Security in Cognitive
  Radio Networks}, 2013.

\bibitem{sha1}
D.~Eastlake, 3rd and P.~Jones.
\newblock {RFC 3174}, {US Secure Hash Algorithm 1 (SHA1)}, 2001.

\bibitem{ganeriwal}
S.~Ganeriwal, C.~P\"{o}pper, S.~{C}apkun, and M.~B. Srivastava.
\newblock Secure time synchronization in sensor networks.
\newblock {\em ACM Trans. Inf. Syst. Secur.}, 11:23--35, July 2008.

\bibitem{hamieh}
A.~Hamieh.
\newblock Powjam: A power reaction system against jamming attacks in wireless
  ad hoc networks.
\newblock WONS 2012, pages 9--15, Jan.

\bibitem{klove}
T.~Klove.
\newblock Error correction codes for the asymmetric channel.
\newblock Technical report, Mathematical Institut University Bergen, 1981.

\bibitem{lazoswisec}
S.~Liu, L.~Lazos, and M.~Krunz.
\newblock Thwarting inside jamming attacks on wireless broadcast
  communications.
\newblock WiSec '11, pages 29--40, 2011.

\bibitem{lazostmc}
S.~Liu, L.~Lazos, and M.~Krunz.
\newblock Thwarting control-channel jamming attacks from inside jammers.
\newblock {\em IEEE Trans. Mob. Comput.}, 11(9):1545--1558, 2012.

\bibitem{liu}
Y.~Liu and P.~Ning.
\newblock Bittrickle: Defending against broadband and high-power reactive
  jamming attacks.
\newblock INFOCOM 2012, pages 909--917, March.

\bibitem{aes}
National and N.~I. S.~T. Technology.
\newblock {\em Announcing the {Advanced} {Encryption} {Standard} ({AES})},
  2001.

\bibitem{phantom}
{Phantom Technologies LTD}.
\newblock {MP806}.
\newblock {\em {http://www.phantom.co.il}}.

\bibitem{Popovski}
P.~Popovski, H.~Yomo, and R.~Prasad.
\newblock Strategies for adaptive frequency hopping in the unlicensed bands.
\newblock {\em Wireless Commun.}, 13(6):60--67, Dec. 2006.

\bibitem{capkun_jsac}
C.~Popper, M.~Strasser, and S.~Capkun.
\newblock Anti-jamming broadcast communication using uncoordinated spread
  spectrum techniques.
\newblock {\em IEEE Journal on Selected Areas in Communications}, 28(5):703
  --715, June 2010.

\bibitem{rappaport}
T.~Rappaport.
\newblock {\em Wireless Communications: Principles and Practice}.
\newblock Prentice Hall PTR, Upper Saddle River, NJ, USA, 2001.

\bibitem{richa}
A.~Richa, C.~Scheideler, S.~Schmid, and J.~Zhang.
\newblock An efficient and fair mac protocol robust to reactive interference.
\newblock In {\em Networking, IEEE/ACM Transactions on}, number~99, pages
  1--12, 2012.

\bibitem{poisel}
P.~A. Richard.
\newblock {\em {Modern Communications Jamming Principles and Techniques (The
  Artech House Information Warfare Library)}}.
\newblock {Artech House Publishers}, Nov. 2003.

\bibitem{Soto}
J.~Soto.
\newblock Randomness testing of the advanced encryption standard candidate
  algorithms.
\newblock In {\em NIST IR 6483, National Institute of Standards and
  Technology}, 2000.

\bibitem{capkun_tosn2010}
M.~Strasser, B.~Danev, and S.~\v{C}apkun.
\newblock Detection of reactive jamming in sensor networks.
\newblock {\em ACM Trans. Sen. Netw.}, 7(2):16:1--16:29, Sept. 2010.

\bibitem{thuente}
D.~J. Thuente, B.~Newlin, and M.~Acharya.
\newblock Jamming vulnerabilities of ieee 802.11e.
\newblock In {\em IEEE MILCOM 2007}, pages 1 --7, Oct. 2007.

\bibitem{blass}
T.~D. Vo-Huu, E.-O. Blass, and G.~Noubir.
\newblock Counter-jamming using mixed mechanical and software interference
  cancellation.
\newblock WiSec '13, pages 31--42, New York, NY, USA, 2013.

\bibitem{trappe}
W.~Xu, W.~Trappe, Y.~Zhang, and T.~Wood.
\newblock The feasibility of launching and detecting jamming attacks in
  wireless networks.
\newblock MobiHoc '05, pages 46--57, New York, NY, USA, 2005. ACM.

\bibitem{xuan}
Y.~Xuan, Y.~Shen, N.~Nguyen, and M.~Thai.
\newblock A trigger identification service for defending reactive jammers in
  wsn.
\newblock {\em Mobile Computing, IEEE Transactions on}, 11(5):793--806, May.

\end{thebibliography}
\end{document}